# Performance of an LPD prototype detector at MHz frame rates under Synchrotron and FEL radiation

**Andreas Koch,**[a,*] **Matthew Hart,**[b] **Tim Nicholls,**[b] **Christian Angelsen,**[b] **John Coughlan,**[b] **Marcus French,**[b] **Steffen Hauf,**[a] **Markus Kuster,**[a] **Jolanta Sztuk-Dambietz,**[a] **Monica Turcato,**[a] **Gabriella A. Carini,**[c] **Matthieu Chollet,**[c] **Sven C. Herrmann,**[c] **Henrik T. Lemke,**[c] **Silke Nelson,**[c] **Sanghoon Song,**[c] **Matt Weaver,**[c] **Diling Zhu,**[c] **Alke Meents,**[d] **and Pontus Fischer,**[d]

[a] *European XFEL GmbH,*
   *Albert-Einstein-Ring 19, 22761 Hamburg, Germany*
[b] *STFC Rutherford Appleton Laboratory,*
   *Didcot, Oxfordshire OX11 0QX, United Kingdom*
[c] *SLAC National Accelerator Laboratory,*
   *Menlo Park, CA, USA*
[d] *DESY,*
   *Notkestraße 85, 22607 Hamburg, Germany*
   *E-mail*: andreas.koch@xfel.eu

ABSTRACT: A MHz frame rate X-ray area detector (LPD – Large Pixel Detector) is under development by the Rutherford Appleton Laboratory for the European XFEL. The detector will have 1 million pixels and allows analogue storage of 512 images taken at 4.5 MHz in the detector front end. The LPD detector has 500 µm thick silicon sensor tiles that are bump bonded to a readout ASIC. The ASIC's preamplifier provides relatively low noise at high speed which results in a high dynamic range of $10^5$ photons over an energy range of 5-20 keV. Small scale prototypes of 32x256 pixels (LPD 2-Tile detector) and 256x256 pixels (LPD supermodule detector) are now available for X-ray tests.

The performance of prototypes of the detector is reported for first tests under synchrotron radiation (PETRA III at DESY) and Free-Electron-Laser radiation (LCLS at SLAC). The initial performance of the detector in terms of signal range and noise, radiation hardness and spatial and temporal response are reported. The main result is that the 4.5 MHz sampling detection chain is reliably working, including the analogue on-chip memory concept. The detector is at least radiation hard up to 5 MGy at 12 keV. In addition the multiple gain concept has been demonstrated over a dynamic range to $10^4$ at 12 keV with a readout noise equivalent to <1 photon rms in its most sensitive mode.

---

[*] Corresponding author.

# Contents

## 1. Introduction

The European X-ray Free-Electron Laser Facility (XFEL.EU) is an X-ray photon source providing laterally coherent X-rays for six experimental stations (start-up configuration) in the range of approximately 250 eV to 25 keV. It is currently under construction and will start user operation in 2016 [1]. The XFEL.EU will provide an X-ray pulse structure with a repetition rate of 4.5 MHz which will not be delivered continuously but within ten pulse trains per second which are separated by 99.4 ms. At the experimental stations each of these trains will be partially populated up to a maximum of 2700 X-ray pulses. The Large Pixel Detector (LPD) development project addresses the particular imaging needs of the Femtosecond X-ray Experiments (FXE) scientific instrument at the XFEL.EU [2]. It will permit time-resolved X-ray absorption, emission, and diffuse-scattering studies (XAS, XES, and XDS) [3]. The LPD detector will be used for liquid scattering and X-ray diffraction experiments where the essential needs are a 1 Mpixel detector with 4.5 MHz frame rate and a high dynamic range for imaging at relatively high X-ray energies (5-20 keV) that allow the detector operation in an air environment. Each of the X-ray pulses may deliver an image that can cover a dynamic range of $10^5$.

      The LPD detector is developed by the Science and Technology Facilities Council at the Rutherford Appleton Laboratory for the XFEL.EU. The project started in 2007 together with

two other detector development projects for the XFEL.EU [4]. First small scale prototypes have been realised in 2011 for electrical tests as well as firmware and software development [5]. Further reduction of noise and improvements in radiation hardness were then achieved with a new Application Specific Integrated Circuit (ASIC) design (ASIC V2). X-ray tests with laboratory sources in 2011 were then followed by first characterisations in May 2013 at the XPP instrument of the LCLS at SLAC (Menlo Park, USA), and at the P11 beamline of PETRA III at DESY (Hamburg, Germany) as reported here.

The LPD key requirements are explained in the next section. An overview of the detector architecture is then given; more details of the front end design can be found in [6]. This article then focusses on the results from the initial beamline tests at SLAC, LCLS and DESY, PETRA III in May 2013.

Table 1: FXE detector requirements and key specifications of Large Pixel Detector LPD.

|  | Requirements FXE | LPD |
|---|---|---|
| **Technology** |  | Hybrid pixel, tiles 4-side buttable |
| **Pixel size** | 500 μm or smaller | 500x500 μm$^2$ |
| **Detector size** | 1kx1k pixel | 1kx1k pixel<br>2-tile prototype: 32x256 pixel<br>supermodule prototype 256x256 pixel |
| **Tiling, hole** | Central, variable hole size | Multiple tiles, 4-side buttable, variable hole (0-10 mm) |
| **Quantum efficiency** | >80% | >80%, 1-13 keV |
| **Fill factor** |  | 86% |
| **Sensor thickness** |  | 500 μm Si |
| **Energy range** | 5-20 keV | 5-20 keV |
| **Dynamic range** | $10^4$ to $10^5$ | $10^5$ at 12 keV |
| **Gain settings** |  | 2 preamplifier feedback capacitances, and 3 parallel gain settings merged off chip, 12 bit ADC |
| **Readout noise** | single photon detection | 0.3 ph rms at 12 keV<br>1000 e$^-$ rms |
| **Frame rate** | 4.5 MHz, 2700 images, 10 pulse trains/s | 4.5 MHz burst mode (analogue on-chip storage of 512 images), 5120 images/s on average |

## 2. Large Pixel Detector (LPD) project

### 2.1 Detector requirements, specifications and architecture

The LPD key requirements are summarised in Table 1. The applications of the FXE instrument at XFEL.EU require a flexible positioning and adjustment of the detector and that combined with the required photon energies drove the choice of silicon sensors and detector operation in air. A central hole of various geometries is also necessary to let the non-diffracted beam pass through the detector to be absorbed further downstream. The hole size is variable within 0-10 mm by a correlated movements of each quarter of the system, more information is given in [5]. The XFEL.EU pulse structure demands a high frame rate of the detector, and this combined with a high dynamic range and low noise is a challenging requirement. In this system, the high frame rate is achieved by a local intermediate analogue storage in the front-end part of the detector. The overall dynamic range is selectable by the choice of two preamplifier feedback settings. The output of the preamplifier is then fed through three parallel gain stages to record the preamplifier output with different precision and signal range, see Figure 1.

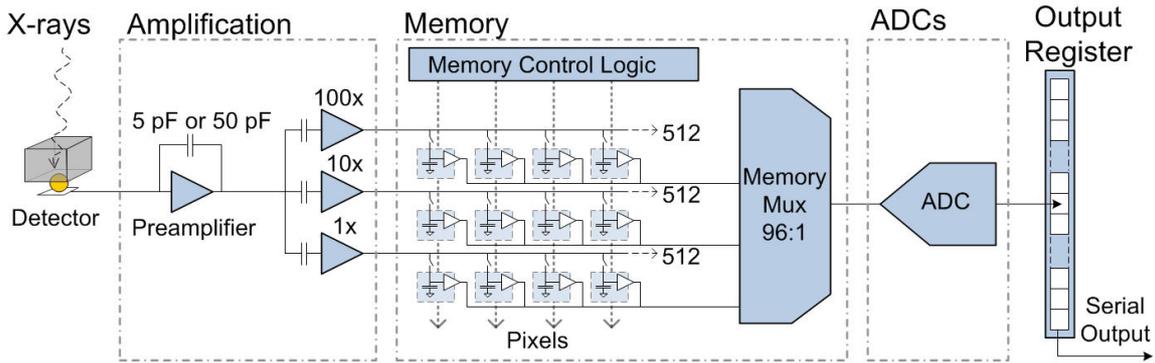

Figure 1: A simplified schematic of the single pixel ASIC architecture.

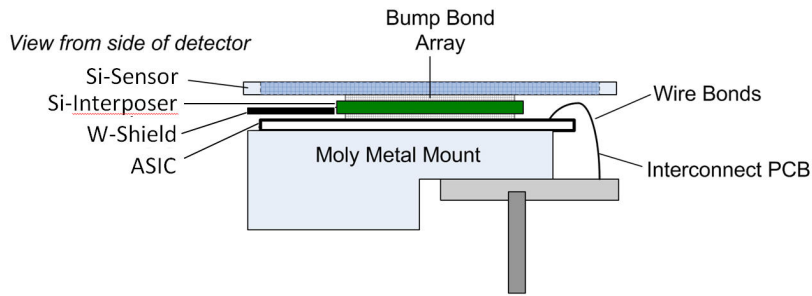

Figure 2: A simplified schematic of a detector tile, not scaling.

The architecture of the detection circuit is shown in Figure 1. In the LPD detector the charge generated by X-rays in the sensor is read out by a custom ASIC into a preamplifier with a selectable 5 pF or 50 pF capacitive feedback (Figure 1). The 50 pF feedback capacitance provides a high dynamic range of $10^5$ 12 keV photons. This amplifier has a dynamic slew correction circuit to allow the reset of such a large capacitance [6]. Following the preamplifier a series of parallel gain stages are implemented. For the smallest signals an amplifier with 100x gain is selected for signal readout. An intermediate gain of 10x is used for mid-range signals and a unit gain amplifier for the largest signal levels. The appropriate gain level for each pixel is selected by the data acquisition system (DAQ) between pulse trains. For applications in which the full dynamic range is not needed, the preamplifier feedback can be switched to 5 pF increasing the signal-to-noise of the system at the expense of dynamic range. Signal and noise performance is discussed below in section 3.2.3 and values are summarised in Table 4. The images are stored within the ASIC in an analogue form at 4.5 MHz during the pulse train. Following the pulse train the analogue data is converted to digital by 16 on-chip ADCs and streamed off the ASIC at 100 MHz via an LVDS output before the following train is recorded. The analogue memory is controlled by a command interface that allows the user to veto and overwrite frames that are not of interest. This makes a better use of the memory available. The command interface is used to drive the ASIC through various states from image capture to data readout.

Each ASIC has 512 pixelated readout channels and eight ASICs are bonded to a sensor tile (32x128 pixels). The sensor tiles are 4-side buttable with a minimum insensitive region between neighbouring tiles equivalent to 4 pixels. Figure 2 shows a side view of a detector tile with a 500 µm thick Si-sensor for X-ray absorption on top of a 600 µm thick Si-interposer for interconnection between sensor and ASIC. Important to note is that the interposer has a smaller area than the sensor that allows X-ray shielding of certain radiation sensitive areas of the ASIC, in particular the memory. A 205 µm thick W-foil is used for this shielding. 16 tiles are then grouped together for a building block, called a supermodule. A supermodule includes all the required readout electronics and interface boards and is the building block for larger area detectors. Control and data acquisition in a supermodule are performed by the custom-designed Front-End Module (FEM) [7]. The FEM is based around a Xilinx Virtex 5 FPGA, which implements flexible firmware and software to control and read out data from the ASICs. Data are transferred from the FEM via 10 Gbps optical Ethernet links for processing and capture downstream. The multi-gain concept in combination with the programmable nature of the FEM allows for a range of operation modes; some examples are given in Table 2.

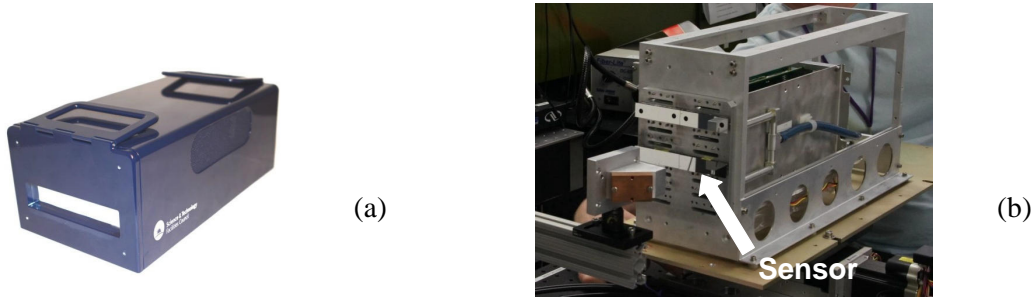

Figure 3: Two LPD detector prototypes are shown. (a) LPD 2-Tile detector prototype with 32x256 pixels, and (b) LPD supermodule detector prototype partly populated with sensor tiles resulting in a 256x256 pixels geometry.

**2.2 LPD prototype systems**

Two types of prototype systems have been realised so far. Firstly, the LPD 2-Tile prototype detector with 32x256 pixels (Figure 3 left) uses all detector front-end components of the final design that determine the signal and noise performance, however, it has no water-cooling circuits which may impact the stability of parameters like offset and gain. Calibration will be therefore limited with this device. However, it is a lightweight prototype of 2 kg and dimensions of 160x120x300 mm$^3$.

Table 2: Operation modes of the LPD.

|  | Mode characteristics |
|---|---|
| Auto gain mode | All three gains are recorded. The DAQ system passes the appropriate data out the system. Default mode of operation. |
| Single gain mode | Only one gain is recorded. |
| 3 gain mode | All three gains are recorded and read. This may slow down the repetition rate. |
| Veto/Trigger mode | Recording long time series by saving only the pulses of interest into memory. |
| Multiple readout mode | Multiple non-destructive readout of memory cells is possible and can be used to test memory cells on signal loss. |
| Long exposure times | Can be used for low signals where the signal of several pulses need to be averaged, or for general characterization with non-FEL like sources. |
| Higher Gain ROI | An arbitrary shaped ROI can be applied to the system to select an alternative dynamic range option. |

In addition the LPD supermodule prototype detector (Figure 3 right) contains 256x256 pixels when fully populated. It uses the same front-end components than the 2-Tile detector. For this prototype cooling is provided; it has a weight of 6 kg and dimensions of 150x250x470 mm$^3$. Both prototypes do not have a central hole for the direct beam.

## 3. Detector characterisation

The LPD prototypes described above have been characterised on laboratory test benches, electrically as well as with X-rays. However, to test the specific features of high intensity, short pulse duration and high repetition rate close to the characteristics of the European XFEL are required. These are only available at other synchrotron or Free-Electron- Laser facilities. For this reason the LPD detector prototypes were tested at LCLS and PETRA III.

### 3.1 Measurements at SLAC / LCLS

Tests at the XPP instrument at SLAC, LCLS (Menlo Park, USA) [8] were undertaken in May 2013 with a monochromatic beam of 9 keV, $10^{10}$ ph/pulse in average, and a repetition rate of 120 Hz. At this facility the signal per pulse fluctuates considerably from pulse to pulse as a result of the SASE process and the use of the monochromator. Therefore, a single shot monitoring of the intensity is necessary for calibration which is provided at the instrument. Some tests were performed using a direct beam with different filters for attenuation, other tests used indirect exposure via the fluorescence of a Cu-block, delivering ~8 keV X-ray radiation uniform over a large area. For the tests at LCLS the LPD supermodule detector was populated with four ASIC V2 tiles. The detector was operated at its nominal rate of 4.5 MHz whereas LCLS delivers pulses at 120 Hz. To enable accurate sampling an LCLS trigger signal was used to synchronise the supermodule's acquisition system with the arrival of the LCLS pulses and apply a tunable delay on the detector side. For each LCLS pulse 10 images at 4.5 MHz were recorded and the delay adjusted so that the X-ray pulse was always recorded in the 4th image.

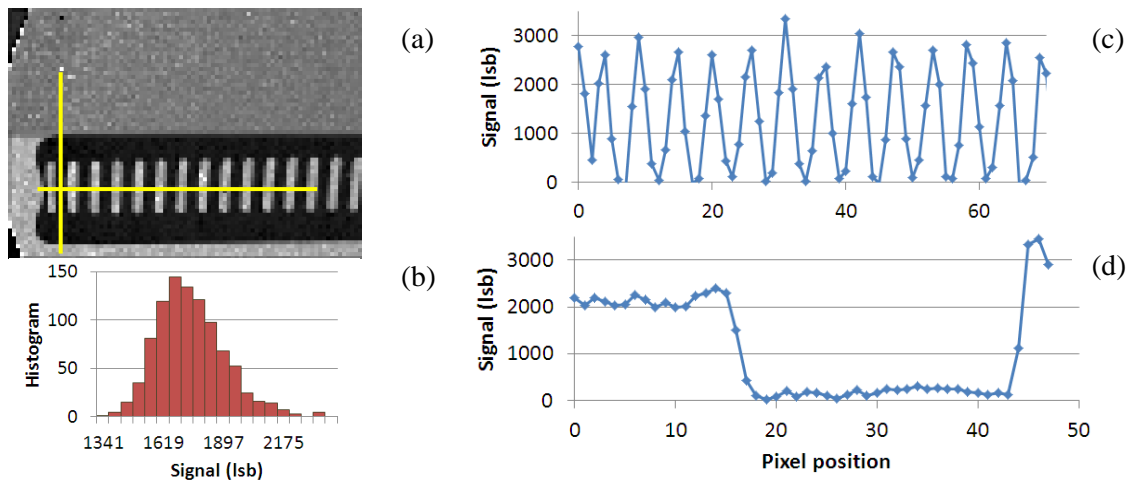

Figure 4: Large area signal exposure on two tiles of the detector. (a) Cu fluorescence image of a Jubilee clip. Yellow lines indicate where the profile plots are taken. (b) Histogram of a uniform ROI in the upper part of the image. (c) Horizontal profile of clip. (d) Vertical profile of clip.

### 3.1.1 Large area signal exposure

Figure 4 shows features of a large area exposure under Cu-fluorescence, in the 5pF, 100x gain mode. The image has had some initial offset and gain correction. In this example the signal intensity is approximately 50 photons per pixel where photons are calculated based on system simulations and first verifications, with an uncertainty of approximately 10%. A precise determination requires a gain calibration, linearity and signal response in different modes which is foreseen to be determined. In Figure 4c and 4d, the Jubilee clip gives an example of the spatial response. Signal variations of approximately 30% between the upper and the lower detector tile are visible and this has been identified as caused by bias voltage stability; improvements are foreseen. To determine the noise, Figure 4b shows a histogram of a uniform region of interest (ROI) in the upper region of Figure 4a. The noise corresponds to the fluctuations of the absorbed X-ray photons.

### 3.1.2 Pixel map of non-responsive pixels

To identify non-responsive pixels in the tested tiles a collimated beam of X-rays from Cu-fluorescence was then scanned across the detector, see Figure 5. In this test the mean signal in the image is 2000 lsb and all pixels with a value below a threshold of one third of the mean signal were assumed to be non-responsive. For LPD the target yield for good pixels is 99.9 – 99.8%, corresponding to 4-8 defective pixels per detector tile. The lower left tile in Figure 5b is within this target value, however, the concentration of defects in the upper left tile is above this and is a result of unconnected pixels (Figure 5b). This higher concentration of unconnected pixels at the edges of sensor tiles is thought to be due to bowing of the sensor during flip chip bonding and we have now developed tools to eliminate this issue in the future. Other types of defects are much rarer and are currently being studied.

### 3.1.3 Signal loss in analogue memory

The analogue memory cells are affected by leakage currents and therefore susceptible to charge loss. To quantify the amount of charge loss we recorded images illuminated with 8 keV Cu-fluorescence. A gain of 100x, 50 pF was used, at a signal of approximately 900 ph/pixel. A multiple readout of the memory cells was then recorded over 20 ms at room temperature. The result is a signal loss rate of 0.7 ph +/-0.3 ph per 100 ms which is a signal level close to the readout noise in the highest gain mode; noise values are given in section 3.2.3. If this result is confirmed for other operation modes, calibration of signal loss of the analogue memory may be required in particular cases. No degradation of this effect by radiation damage is expected up to 25 keV since the memory cells are shielded by the 205 μm thick tungsten sheets.

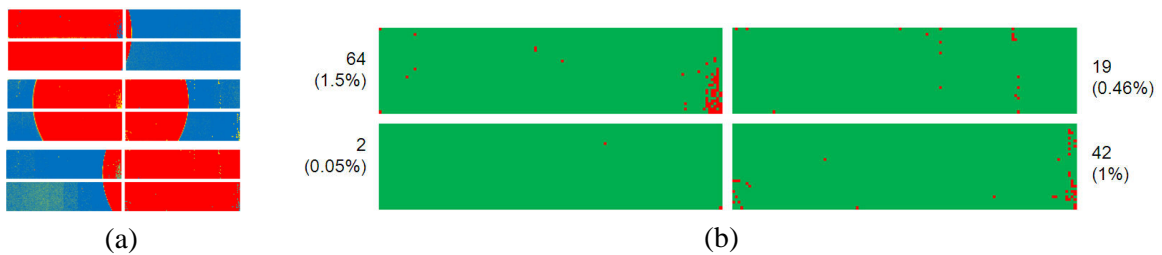

Figure 5: Pixel map of non-responsive pixels. (a) Scan of beam (in red) across tiles in three steps. (b) Defective pixels are shown in red. The number of defects per tile and its percentage value are given.

## 3.2 Measurements at DESY / PETRA III

Tests at the P11 beamline at DESY, PETRA III were also done in May 2013 with a monochromatic beam of 12 keV, in 40 bunch mode at a pulse rate of 5.2 MHz. The intensity was measured to be stable below a few per cent at $2\times10^{13}$ ph/s with a beamsize of the defocused beam of 2x3 mm$^2$. For the tests at PETRA III the LPD 2-Tile detector was used with two ASIC V2 tiles. The detector was synchronised to the PETRA III bunch clock of 5.2 MHz to simplify operation and no performance losses by operation at this higher speed were observed.

### 3.2.1 Radiation hardness

Radiation hardness was probed on several test regions of a detector tile at different entrance dose levels (see Figure 2 for the architecture of a detector tile).
The location of the following regions for the radiation tests is shown in Figure 6:

1. Exposure on sensor, interposer, ASIC pixels and ASIC bias circuit, 5 MGy.
2. Exposure on sensor, interposer, ASIC pixels and ASIC bias circuit, 50 MGy.
3. Exposure on sensor, metal shield and ASIC memory, 20 MGy.

The beam of 2x3 mm$^2$ was scanned in steps across the pre-defined areas with exposures of minutes to hours to achieve the required doses. At 4.5 MHz readout only at a dose of 50 MGy are any damage effects visible (see Figure 6a). The amplifiers in the region highlighted now do no longer function correctly. By increasing the integration time by two orders of magnitude the change in leakage current in the sensor as a result of the other exposures becomes visible; this is shown in Figure 6b. Here, the regions that accumulated 5 and 20 MGy show a higher increase in leakage current than the 50 MGy region. This effect is presently under study and not yet fully understood. The maximum leakage current recorded is 1 nA/pixel which is a factor of ten higher than for unexposed pixels (Table 3). The noise of this level of leakage current is a factor of ten lower than the other contributions to the readout noise. For the 50 MGy exposure (region 2 in Figure 6a and 6b), the affected ASIC region appears displaced since an ASIC pixel is not exactly below the corresponding sensor pixel (see also Figure 2). At higher photon energies the ASIC is more susceptible to radiation damage since the sensor is less absorbing. For example, at 20 keV only approximately 1 MGy instead of 50 MGy entrance dose is required to deposit in the ASIC in region 2 a similar energy. In addition, as expected, no degradation of the memory has been observed because of the radiation shielding.

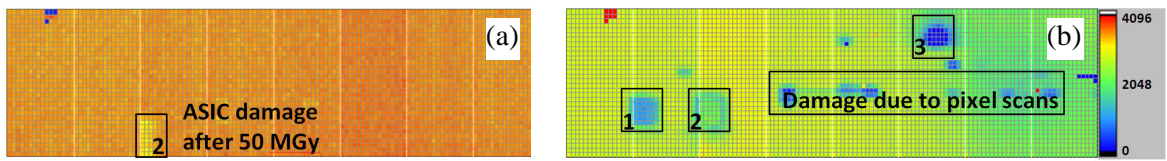

Figure 6: Radiation hardness tests at different regions on a detector tile of 32x128 pixels. (a) Dark image at 4.5 MHz readout, 120 ns integration time . (b) Leakage current: 100 times longer integration time.

Table 3: Leakage current. Measured values in nA/pixel, calculated values in photons.

|  | Not exposed | 5 MGy | 50 MGy |
|---|---|---|---|
| Leakage (nA/pixel) | < 0.10 | 1.1 | 1.0 |
| Leakage (12 keV ph/pixel/220 ns) | 0.042 | 0.47 | 0.42 |

More detailed investigations are presently on-going which will reveal in more detail up to what limits calibration can compensate for these effects. Additionally, annealing effects will be studied.

### 3.2.2 Dynamic range

At Petra III a direct, defocused X-ray beam was used to probe part of the signal range of the detector. Using Al-filters of different thicknesses the signal was varied between 5 ph/pixel in the highest gain mode and up to $5\times10^4$ ph/pixel in the lowest gain mode. In addition the detector entrance signal was measured with a Si-photodiode. The characterisation and study of saturation and linearity in each gain stage is on-going and will be reported in the future but all the effects seen are within expectations.

### 3.2.3 Readout noise

This test has been done at system level at the beamline in order to have a realistic environment since DAQ, cabling and electromagnetic interference (EMI) may influence the electronic noise of the system. The only difference was that an external power supply for the detector was used since the internal power card was not fully optimised at this stage. All the noise values in Table 4 are again referred to the input of the detector in absorbed photons of 12 keV (see remark in section 3.1.1). The results for the system noise in Table 4, column (3) refer to an image corrected for offset, gain and defects and the uncertainties are <3%.

The measured noise values for the 2-Tile detector in the beamline environment, Table 4, column (3) correspond well to the noise measured with a single chip test system, column (2), except for gain 1x. According to design simulations, column (1), a lower noise level was expected, especially in the high gain modes. Further studies to understand these discrepancies are on-going and may stem from inaccuracies in process models and the simulation test-benches.

Table 4: Readout noise and maximum signal at different amplifier gain settings. Data in each gain stage is digitised by a 12 bit ADC.

| Gain | | Maximum signal (12 keV photons) | Readout noise rms in 12 keV photons | | |
|---|---|---|---|---|---|
| | | | ASIC V2 simulation (1) | ASIC V2 measured (2) | System measured with ASIC V2 (3) |
| 50 pF | 1x | 100000 | 100 | 107 | 160 |
| | 10x | 10000 | 11,0 | 14,0 | 13,6 |
| | 100x | 1000 | 1,9 | 4,1 | 3,9 |
| 5 pF | 1x | 10000 | 10,0 | 11,9 | 24,5 |
| | 10x | 1000 | 1,6 | 1,6 | 1,6 |
| | 100x | 100 | 0,30 | 0,46 | 0,56 |

## 4. Conclusions

The first tests of an LPD 2-Tile and supermodule system under synchrotron and FEL radiation demonstrate that the LPD concept of imaging frame rates of up to 4.5 MHz combined with analogue on-chip storage and multiple gain stages to cover a high dynamic range of $10^5$ for 12 keV photons is working well. The operating conditions for these tests have been close to those of the European XFEL.

The new ASIC V2 design provides detector radiation hardness and a readout noise level within or close to the specifications. These two parameters are critical for scientific applications at the European XFEL. Radiation hardness now is on a sufficient level to start operation although it needs more understanding and characterisation. Other parameters like the memory signal loss, the signal range, and pixel defects have been tested under typical conditions and mostly are within specifications. Optimisation of the system configuration and bias conditions is on-going to further improve the system performance. This includes improvements of electronic noise and number of pixel defects. Detailed characterisation and calibration over the whole range of operating conditions of the detector in terms of photon energy and signal intensity at the various detector gain settings will be the next step.

## Acknowledgments

We would like to thank Christian Bressler and his group for continuous valuable discussions during the technical optimisation of the LPD detector system.

Portions of these tests were carried out at the light source PETRA III at DESY, a member of the Helmholtz Association (HGF). We would like to thank Jan Meyer and Jan Roever for assistance in using beamline P11. Other tests were carried out at the Linac Coherent Light Source (LCLS) at the SLAC National Accelerator Laboratory. LCLS is an Office of Science User Facility operated for the U.S. Department of Energy Office of Science by Stanford University. We would like to thank Chris Kenney and Aymeric Robert. Without their support the LPD beamline tests at LCLS would not have been possible.

Further we would like to thank all members of the XFEL.EU Detector Development team and the XFEL.EU DAQ & Control Systems team who supported these tests.